\documentclass[12pt,twoside]{article}
\title{Most recent changepoint detection in Panel data} 
\author{Lawrence Bardwell, Idris Eckley, Paul Fearnhead, Simon Smith \& Martin Spott}
\date{\today}  

\usepackage[left = 2.5cm, right = 2.5cm, top = 2cm, bottom = 2.5cm]{geometry} 
\usepackage{setspace}
\usepackage{graphicx}
\usepackage{amsfonts} 
\usepackage{amsmath}
\usepackage{float}
\restylefloat{table}
\usepackage{natbib} 
\usepackage{placeins}
\usepackage{caption}
\usepackage{subcaption} 
\usepackage{amsthm}
\numberwithin{equation}{section}
\usepackage[ruled]{algorithm2e}
\usepackage{titlepic}
\DeclareMathAlphabet{\mathpzc}{OT1}{pzc}{m}{it}

\usepackage{color}
\usepackage[toc,page]{appendix}
\usepackage{fancyhdr} 
\usepackage{url}
\usepackage{xcolor}
\usepackage{bbm}
\usepackage{verbatim}
\usepackage{multirow}
\usepackage{changebar}

\DeclareMathOperator*{\argmin}{argmin}

\usepackage[pdftex]{hyperref}  
\graphicspath{{Figures/}}

\begin{document}
\bibliographystyle{apalike}  
\maketitle	
\begin{abstract}
Detecting recent changepoints in time-series can be important for short-term prediction, as we can then base predictions just on the data since the changepoint. In many applications we have panel data, consisting
of many related univariate time-series. We present a novel approach to detect sets of most recent changepoints in such panel data which aims to pool information across time-series, 
so that we preferentially infer a most recent change at the same time-point in multiple series. Our approach is computationally efficient as it involves analysing each time-series independently to obtain a profile-likelihood 
like quantity that summarises the evidence for the series having either no change or a specific value for its most recent changepoint. We then post-process this output from each time-series to obtain
a potentially small set of times for the most recent changepoints, and, for each time, the set of series which has their most recent changepoint at that time.  
We demonstrate the usefulness of this method on two data sets: forecasting events in a telecommunications network and 
inference about changes in the net asset ratio for a panel of US firms.

\end{abstract}
{\bf Keywords:} Breakpoints, Changepoints, Forecasting, Panel data, Structural Breaks.

%
\section{Introduction}
\label{sec:intro}

There are many modern applications where high-dimensional observations are collected and stored over time. This type of data can be viewed as a (potentially large) collection of time series and in the literature is often known as panel data. For an overview of this area see \citet{panel_data}.

We are interested in structural changes, also known as changepoint detection.  
For an overview of some of the methods used on univariate time series see \citet{JTSA:JTSA12035}. 
In this work, however, we will look at structural changes in panel data. 
Some recent work in this area includes \citet{doi:10.1080/01621459.2014.957545}, \cite{ma2016pairwise} and \citet{doi:10.1080/01621459.2014.920613}. Applications of these methods to detect changes 
occur in many areas such as finance, bioinformatics and signal processing \cite[]{RSSB:RSSB12079,NIPS2010_4157,cao2015changepoint}.

Our work is motivated by a real-life problem of predicting the number of events that occur across a telecommunications network.
We have weekly data on the number of events in the network, with this number recorded for each of a set of event types and for each of a set of geographical regions.
Being able to make short-term predictions of future event counts is important for planning. These event counts are observed to change over time, often abruptly, and it 
is  natural to model the time series data using a changepoint model. 

\begin{figure}[t!]
  \centering
  \includegraphics[scale=0.4]{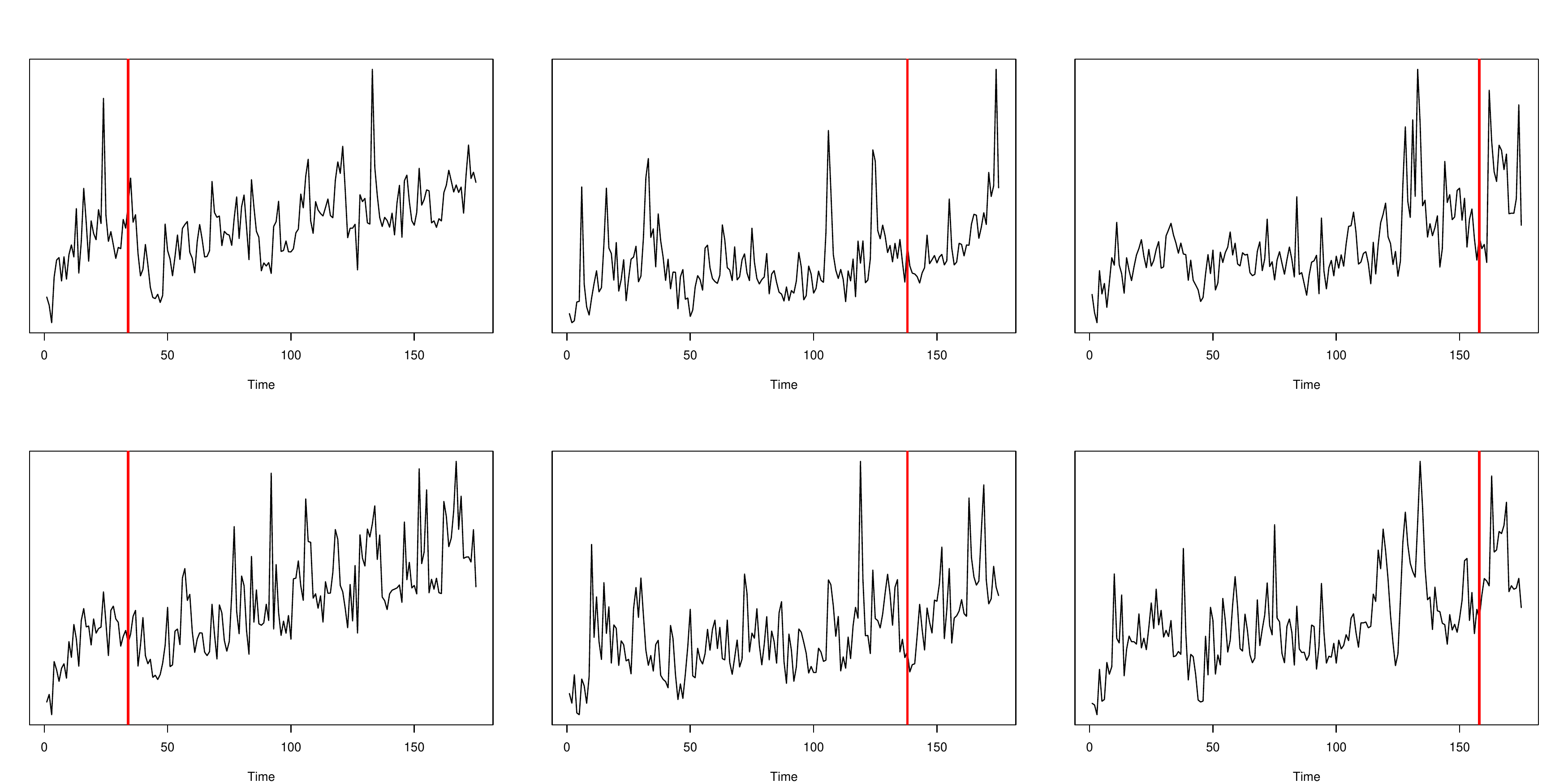}
  \caption{An example of six of the event count time series. These show different patterns. The left-hand column has two series consistent with a constant positive trend since around week 40. The middle column show series with evidence for a recent increase in trend around week 140. The right-hand column shows series with evidence for a decrease in the rate of events from around week 160. In each case we show our estimate of the most recent changepoint -- see Section \ref{sec:BT} for more detail.
  }
  \label{Fig:1}
\end{figure}

The challenge with analysing the data is dealing with the large number of separate time series, one for each event type and region pair. In total there are 160 time series. Six example time series are shown in Figure \ref{Fig:1}. It is natural to
assume that some reasons, such as large external factors, that affect the event count for one time series may also affect the event counts for other time series. However, not all time series
may see a changepoint at exactly the same time. We would like a changepoint method that has the flexibility to encapsulate, but does not force, time series to share
common changepoints. As our primary interest is in short-term prediction, we particularly want a method that is accurate in estimating the location of the most recent change-point for each time series, so that
we can use the data since that change-point to predict the likely number of events in the future.

Detecting changepoints in multiple time series introduces computational challenges that are not present when analysing a single time series. A simplistic approach to the problem would thus be to
try and apply univariate changepoint methods \cite[]{JTSA:JTSA12035}. There are two ways of doing this. One is to analyse each time series separately. The other is to aggregate the time series, and
analyse the resulting univariate series. Each method has its drawbacks. The former will lose power when detecting changepoints, as it ignores the information that different time series are likely to have
changepoints at similar times. The latter approach can perform poorly if the signal from changepoints that affect a small number of series is  swamped by the noise in the remaining series when they are aggregated.

An alternative approach to analysing data of this form is to treat the data as a single time series with multivariate observations. We then model the multivariate data within a segment, and allow for this
model to change, in an appropriate way, between segments. This approach is taken by \citet{lavielle}, who model data as multivariate Gaussian but with a mean that can change from segment to segment. Similarly,
\citet{ecp} present a non-parametric approach to detecting multiple changes in multivariate data. However, like aggregating the data, these methods may lack power if a change only affects a small proportion of the
time series. \cite[Though see][for ideas that try to overcome this problem]{wang2016high}.


Recently there have been methods specifically designed for detecting changes that affect only a subset of series. \citet{RSSB:RSSB12079} and \citet{DCBS} propose a way to detect a single,
potentially common, changepoint in such data. They consider a novel, non-linear, way of combining summaries of individual time series, so-called CUSUM statistics, 
that contain information about the presence and location of a changepoint. 
The intuition is to retain CUSUM values from all series that show strong evidence for a change at a given time-point, but down-weight the values from other time series. 
Thus they are able to share information across time series without any signal being swamped by noise from series which do not share the common changepoint.  
Similarly, \citet{xie2013} introduce a generalised likelihood ratio test for detecting a single common changepoint that affects only a subset of series. This test needs an estimate of the proportion of series affected by the change, and this estimate then affects the weight given to evidence for
a change from each series. Again, the intuition of the approach is to give large weight to series that show strong evidence for a change, but lower weight to those with little evidence.

The approaches of \citet{RSSB:RSSB12079}, \citet{DCBS} and \citet{xie2013}  can each be used within a binary segmentation procedure to find multiple changes. 
Empirical results in these papers
show that this type of approach can be more powerful than either analysing series individually or aggregating them. 

As we are primarily interested in estimating the most recent changepoint for each time series, we take a different approach.  Our approach is focussed primarily on detecting the
most recent changepoint in each time series. It does this by partioning the panel of time series into groups each of which share the same most recent changepoint, with, potentially, a group
corresponding to time series with no change.
%
This is achieved by analysing each time series independently using a penalised cost, or penalised likelihood, approach to detecting changes \citep{Lavielle20051501,pelt,FPOP}. From each analysis we output
a measure of evidence for the most recent changepoint being at each possible time-point, or that the series has no change. 
 We then post-process the output from these analyses in a way that encourages time series to share a common most recent change. This post-processing step involves trying to partition the time series
 into a small number, $K$, of groups that share the same value for their most recent changepoint.
 We show that this post-processing step can be
formulated in terms of solving  a combinatorial optimisation problem, known as the $K$-median problem. Whilst this problem is NP-hard, we use a heuristic solver that is computationally inexpensive,
and, empirically, works well in terms of the estimated most recent changepoints.


The outline of the paper is as follows.
Firstly we define the problem of finding the most recent changepoint in a univariate time series using a penalised cost approach, and show how this can be extended to panel data.
To infer the most recent changepoints requires solving a combinatorial optimisation problem. We discuss how to solve this in Section \ref{sec:mmrc}. In Section \ref{sec:sim_study} we evaluate our method, 
and compare it with a number of alternatives 
on simulated data.
We then apply our method to two real data applications. The first data set represents a telecommunications event time series, shown in Figure \ref{Fig:1}, where 
the aim is for improved prediction. Secondly, we analyse financial data from a large number of firms. In this
application we are more concerned about detecting the locations of most recent changepoints and the sets of firms that change. 
The aim of this is to understand the causes of these changes, for example whether they be legal changes that affect specific sectors, or wider economic changes. 
Finally we end with a discussion on the advantages and limitations of our method. 

\section{A Penalised Cost Approach to Most Recent Changepoint Detection}
\label{sec:MRC}

We begin by assuming we have panel data consisting of $N$ time series of length $n$. Denote the $i$th time series by $y_{1,i},\ldots,y_{n,i}$. Throughout we will use the notation $y_{s:t,i}$ to denote the subset of
observation from time $s$ to time $t$ inclusive. 

Our approach to detecting the common most recent changepoints is based on a penalised cost approach. We will first describe how this approach can be used to analyse individual time series, before then
explaining how the output from these individual analyses can be combined to estimate a set of common most recent changepoints for our $N$ series.

\subsection{Analysing a Univariate Time Series}
\label{sec:mrc_meth}

First consider analysing data from one of the $N$ time series in our panel data. To simplify notation we will drop the subscript that denotes which time series, and instead denote the data by $y_{1:n}$. We
will denote the number and position of changes by $m$ and $\boldsymbol\tau=(\tau_1,\ldots,\tau_m)$ respectively. We assume the changepoints are ordered, and define $\tau_0=0$ and $\tau_{m+1}=n$.


A penalised cost approach to detecting changepoints in this time series involves introducing a cost associated with each putative segment. This cost is often derived by modelling the data within a segment, and defining
the cost to be proportional to minus the maximum likelihood value for fitting that model to a segment of data. If our model for data in a segment is that they are IID with some density $f(y|\theta)$, where $\theta$ is
a segment-specific parameter, then we can define the cost for a segment $y_{s:t}$ as
\[
 \mathcal{C}(y_{s:t})=-2\max_\theta \sum_{u=s}^t \log f(y_u|\theta).
\]
The segment cost function can include a component that depends on the length of segment as is used in some penalised cost approaches \cite[]{davis2006structural,Zhang2007}.

To make this idea concrete we will give two examples of cost functions that we will use later. The first is for detecting a change in mean. A simple model is that the data in a segment is IID Gaussian with common
known variance, $\sigma^2$, and segment specific mean, $\theta$. In this case we get
\[
 \mathcal{C}(y_{s:t})=-2\max_\theta \frac{-1}{2\sigma^2}\sum_{u=s}^t\left(y_u-\theta\right)^2=\frac{1}{\sigma^2}\sum_{u=s}^t\left(y_u-\frac{\sum_{v=s}^ty_v}{t-s+1}\right)^2.
\]

The second is where we model the mean of the data within a segment as a linear function of time, but allow this linear model to vary between segments. Denote $\theta=(\theta_1,\theta_2)$ to be the segment intercept and
slope. If the noise for this model is IID Gaussian we then get
a segment cost
\[
 \mathcal{C}(y_{s:t})=\frac{1}{\sigma^2} \max_\theta \sum_{u=s}^t \left(y_u - \theta_1 - u\theta_2 \right)^2.
\]
We use this model for analysing the data presented in the introduction, however in that application some time series have clear outliers. To make our inferences robust to these outliers we follow \citet{Fearnhead/Rigaill:2016}
and instead use a segment cost
\begin{equation} \label{eq:Robust}
 \mathcal{C}(y_{s:t})=\frac{1}{\sigma^2} \max_\theta \sum_{i=s}^t \min\left\{\left(y_i - \theta_1 - i\theta_2 \right)^2,4\sigma^2 \right\}.
\end{equation}
This cost limits the impact of outliers if their residuals are greater than $2$ standard deviations away from the segment mean.

For all these costs we require knowledge of $\sigma^2$, the residual variance (or in the latter example, the variance of the non-outlier residuals). In practice we use a simple 
and robust estimator of $\sigma$, based on the median absolute deviation of the differenced time series \cite[]{Fryzlewicz2012}.

Once we have defined a segment cost, we then define a cost for a segmentation as the sum of the segment costs for that segmentation. To segment the data, and find the changepoints, we then want to minimise
this cost over all segmentations. However to avoid over-fitting we add a penalty, $\beta>0$, for each segment.
Thus to segment the data  we solve the following optimisation problem 
\begin{align}
\label{eq:full_PELT}
  \min_{m,\boldsymbol\tau} \sum_{j=1}^{m+1} \left[ \mathcal{C}(y_{(\tau_{j-1}+1):\tau_j}) + \beta \right].
\end{align}

The choice of $\beta$ in this approach is important. Higher values for $\beta$ will mean fewer changepoints detected. There are various suggestions for how to choose $\beta$, and the most
common for detecting changes in a single time series is the BIC criteria. If
our segment specific parameter is of dimension $p$, then this corresponds to $\beta=(p+1)\log n$.
This has good theoretical properties, if our modelling assumptions are correct 
\cite[e.g.][]{yao1987}. However care is needed in practice where this is not the case, see \citet{CROPS} for guidance in selecting 
an optimal value for $\beta$ for a given a time series.  


Solving \eqref{eq:full_PELT} is possible using dynamic programming.
This requires the solution of a set of intermediate problems. Define $F(t)$ for $t=1,2,\hdots , n$ as 
\begin{align}
\label{eq:PELT_optim}
  F(t) = \min_{ \boldsymbol\tau } \left\{ \sum_{j=1}^{m+1} \left[ \mathcal{C}(y_{(\tau_{j-1}+1):\tau_j}) + \beta \right] \right\},
\end{align}
where the minimisation is over $m$ and $0=\tau_0<\tau_1<\cdots<\tau_m<\tau_{m+1}=t$.
Thus $F(t)$ is the minimum cost for segmenting data $y_{1:t}$.
The functions $F(\cdot)$ can be efficiently calculated, for example using the PELT \cite[]{pelt} or FPOP \cite[]{FPOP} algorithms, as
\[
 F(t)=\min_{s<t} \left\{F(s)+\mathcal{C}(y_{s+1:t})+\beta\right\}.
\]

Recalling that our interest is in detecting the most recent changepoint, let us consider $G(r)$, 
which we define to be the minimum cost of the data conditional on the most recent changepoint prior to $n$ being at time $r$. 
This is related to $F(r)$ as it is just the minimum cost of segmenting  $y_{1:r}$ plus the cost of adding a changepoint and the cost for segment $y_{(r+1):n}$,
\begin{align}
\label{eq:prof_lik}
  G(r) = F(r) + \mathcal{C}(y_{(r+1):n}) + \beta, \mbox{ for $r=1,\ldots,n-1$,}
\end{align}
with $G(0)=\mathcal{C}(y_{1:n})$. This quantity can be viewed as related to the idea of a profile likelihood, as we have optimised over all nuisance parameters 
(the number and locations of the changepoints prior to the most recent changepoint). 
It is trivial to see that our estimate for the most recent changepoint is given by $\argmin_{r \in \{0, \hdots , n-1 \}} G(r)$. If the most recent changepoint is at $r=0$, then this corresponds to
no change within the time series.


\subsection{Extension to panel data}
\label{sec:panel_data}


We now return to the problem of finding a set of common most recent changepoints in our panel data. Let $G_i(r)$ denote the minimum cost  for segmenting series $i$ with a most-recent changepoint at $r$, defined in
(\ref{eq:prof_lik}).
Our idea is to search for a set of $K$ locations for the common most recent changepoints for our $N$ series. 

Firstly assume that an appropriate value for $K$ is known. Denote a set of common most recent changepoints as
 $\mathbf{r}_{1:K} = (r_1 ,  \hdots , r_K)$. 
 For the $k$th most recent changepoint, located at $r_k$, then there will exist a set, $I_k \subset \{ 1, 2 , \hdots , N \}$, 
 such that all series $i \in I_k$ the most recent changepoint is located at $r_k$. The sets $I_{1:K}$ will partition the full set of series $\{1,2, \hdots ,N\}$.


It is natural to estimate the $\mathbf{r}_{1:K}$, and the associated sets, by the values that minimise the sum of costs for each series
\begin{align} 
\label{eq:mmrc}
C_K  =  \min_{I_1, \hdots , I_K \hspace{5pt}  m_1,\hdots,m_K  } \sum_{k=1}^{K}  \sum_{i \in I_k} G_i(r_k).  
\end{align}
The minimisation of \eqref{eq:mmrc} is challenging, however we will describe a method adopted from the field of combinatorial optimisation to solve it for a given value of $K$  in Section \ref{sec:mmrc}. 

In practice we do not know what value of $K$ to choose. Thus to choose $K$ we resort to minimising a penalised version of (\ref{eq:mmrc}). We first solve the optimisation problem in (\ref{eq:mmrc}) for a range of $K$,
and then choose the value of $K$ that minimises
\[
 C_K+ N \log_2 K + K \log_2 n,
\]
where $\log_2$ is log base two. This uses a minimum description length criteria \cite[]{grunwald2007minimum}, and the penalty can be viewed as the log, in base two, of the model complexity for allowing $K$
most recent changepoints: the number of choices of the $K$ changepoints is approximately $n^K$ and then each of the $N$ time series can choose which of the $K$ most recent changepoints to have, which gives $K^N$ possible choices.

This approach penalises adding most recent changepoints. Thus when we implement our method we use a value of $\beta$, the penalty for adding a change used in calculating $G_i(r)$, which is slightly lower
than the BIC choice. Specifically, we suggest using $\beta=(p+1/2) \log n$, as on simulated data with no change, values of $\beta$ lower than this produce $G(r)$ functions that on
average get smaller as $r$ increases for $r\geq 1$ -- which suggests smaller choices of $\beta$ would be biased towards adding erroneous very recent changepoints. By comparison our choice of $\beta$ produced $G(r)$ functions whose average
value appeared constant for $r\geq 1$.




\section{Optimal set of most recent changepoints}
\label{sec:mmrc}

We now turn to solving the optimisation problem in \eqref{eq:mmrc} for a fixed value of $K$. 
Solving this is computationally challenging if a brute force method is applied, due to the exponentially large number of ways of choosing either $\mathbf{r}_{1:K}$ or the sets $I_{1:K}$. 
However it can be reduced to a well studied problem in the field of combinatorial optimisation. 

To formulate this problem we proceed as follows. Let $\mathbf{G}$ be a matrix of the 
conditional costs that we defined in \eqref{eq:prof_lik}, so that $\mathbf{G}_{ir} = G_i(r)$ which is the optimal cost of the most recent changepoint being at time $r$ in the $i$th series, where $i=1,2,\hdots,N$ 
and $r=0,1,\hdots,n-1$. We want to find the $K$ columns of $\mathbf{G}$ such that if, for each row, we take the minimum of elements in these columns, and then sum these across all  $N$ rows,  
the total is minimised. This allocates each of the $N$ series into $K$ disjoint classes according to which series are affected by a specific most recent changepoint. The specific optimisation problem is
\begin{align*}
\min_S \sum_{i=1}^{N} \min_{r \in S} \mathbf{G}_{ir}, \textrm{ where $S \subset \{0,1,\hdots ,n-1\}$ and $\lvert S \rvert = K$}.
\end{align*}
It turns out that this optimisation problem is mathematically equivalent to the so-called $K$-median problem \cite[]{NET:NET20128}. This problem can be formulated, 
and solved, as an integer program  with binary variables $x_{ir}$ and $z_r$ where
\begin{align*}
  x_{ir} =
\left\{
	\begin{array}{ll}
		1  & \mbox{if series $i$ has most recent changepoint at time $r$} \\
		0 & \mbox{otherwise, }
	\end{array}
\right.
\end{align*}
and 
\begin{align*}
  z_{r} =
\left\{
	\begin{array}{ll}
		1  & \mbox{if there is a most recent changepoint in any series at time $r$} \\
		0 & \mbox{otherwise. }
	\end{array}
\right.
\end{align*}

The objective is then simply to solve the following problem:
\begin{align}
  \min \sum_{i=1}^{N}\sum_{r=0}^{n-1} &\mathbf{G}_{ir}x_{ir} \label{eq:objective} \\
\text{subject to } \hspace{30pt}  \sum_{r=0}^{n-1} x_{ir} &= 1, \forall i, \label{eq:rows} \\
x_{ir} \leq  z_{r}&,\forall i,r, \label{eq:cols1} \\
\sum_{r=0}^{n-1} z_r &= K. \label{eq:cols2}
\end{align}
Here constraint \eqref{eq:rows} ensures each series has only one most recent changepoint, whilst the two remaining constraints, \eqref{eq:cols1} and \eqref{eq:cols2},  ensure that $K$ different most recent changepoints are selected. 

Approaches for solving the $K$-median problem are discussed in \citet{NET:NET20128} and references therein.
We use the method of \cite{teitz1968heuristic}, available within the {\texttt{R}} package {\texttt{tbart}}. This is a simple algorithm that tries to improve on a current solution by replacing one of the
$K$ values for a most recent changepoint with a value that is not currently in the set of most recent changepoints. It loops over all such pairs, and makes the replacement if it will reduce the objective function
(\ref{eq:objective}). This is repeated until there is no replacement that will improve the objective any further. This method is heuristic, in that it is not guaranteed to find the global optimum to the optimisation
problem. However we found that it is computationally efficient and empirically leads to
good estimates of the most recent changepoints, as shown below.

\section{Simulation study}
\label{sec:sim_study}


As described in the introduction, there are a number of methods in the literature that allow us to detect multiple changes in panel data. We compare our method, which we call MRC, 
to several of these to see empirically how they compare. None of these alternative methods were specifically designed to just estimate the most recent changepoints, and we are unaware of any other
methods that focus solely on this. Furthermore some of these methods are able to infer quantities, such as earlier common changepoints, that MRC cannot.

The alternative methods can be split into two groups. The first set of methods estimate common changepoints for each series. We compare with three such approaches. These are analysing the aggregated data (AGG) and
two approaches for detecting common changepoints in multivariate data. The latter two methods are the approach of  \citet{lavielle} which models data within a segment as multivariate Gaussian with known covariance (MV); and  the ECP method \cite[]{ecp}, which is a non-parametric changepoint detection procedure (ECP).

Both the AGG and MV methods require a choice of penalty and we use the BIC penalty. However, for the ECP method every proposed changepoint is tested for statistical significance using a permutation test and a threshold obtained via a bootstrap which is described in \citet{ecp}.

The second group of alternative methods includes two methods that can estimate common changepoints that affect only a subset of the time series. 
The simplest method we consider (IND) involves analysing each series in the panel
independently  and finding the most recent changepoint in each series. The second method in this group is Double CUSUM Binary Segmentation (DCBS) \cite[]{DCBS}. Whilst the focus of this method, and the theory
that underpins it, is on consistently detecting the location changes, the paper also mentions an intuitively natural way of identifying which subset of series changes at each changepoint.
%
We again use the BIC penalty when segmenting each series as part of the IND method. The DCBS method has two parameters that need to be chosen. The first parameter, $\psi$, is related to the expected degree of sparsity or the number of series affected by a change compared to the total number of series. Guidance is available on how to choose this parameter in \citet{DCBS}. 
The second parameter, $\pi^{\psi}$, is the threshold for testing whether or not a change is significant as is done in the ECP method mentioned above. This threshold is chosen using a bootstrap style procedure where the null hypothesis of no changepoint is assumed and some empirical quantile of this distribution is taken. We chose this parameter by simulating 100 replications from the null hypothesis, i.e.\ no changepoints at all, and measured the proportion of false positives for a number of different values for $\pi^{\psi}$. In practice, we found that a value of $\pi^{\psi} = 10$ worked well.

Each panel data set we simulate consists of 100 series all having length 500. For a given value of $K$  we first simulate $K$ distinct values for the most recent changepoints from the set $\{300,320,\ldots,480\}$. This ensures each most recent changepoint position is at least 20 time-points away from all other positions, which helps interpretation when
we measure the accuracy of methods in detecting the location of the changes. We partition our 100 time series evenly across the $K$ most recent changepoint locations.
 We then simulate earlier common changepoints by first simulating potential changepoints independently with probability 0.02
at each time-point prior to the earliest most recent common changepoint. For each of these we simulate a probability from a uniform distribution, and then simulate that a changepoint appears in each time series independently
with this probability.  The observations in each of the segments are IID Gaussian distributed with mean $\mu$ drawn from its prior distribution $\mathcal{N}(0,2^2)$. 
For simplicity we keep a fixed variance $\sigma^2 = 1$ for all the observations. In this study the parameter of the last segment differs by $\epsilon$ from the mean in the penultimate segment, with the sign 
of the change being chosen uniformly at random for each time series.
We use $\epsilon = 1$ for the studies in Cases 1, 3 and 4 below,  whereas for Case 2 we look at the effect of varying $\epsilon$. 



In the first three studies we consider the accuracy of estimates of the most recent changepoints and which series are affected. We only compare 
IND and DCBS with MRC, as these are the only methods that estimate which series are affected by each of the most recent changepoints. 
%
We evaluate these three methods on a number of different criteria. 
A specific changepoint is then defined as being detected if it is within 5 time points of an estimated changepoint, and we calculate the proportion of changepoints that are detected. 
To define the location accuracy we take only those changepoints that are detected then 
take the average of the absolute difference between the true and estimated locations. 

Two of the methods we consider, namely MRC and DCBS, return more information than IND, including the estimated number of most recent changepoints 
$\hat{K}$ and the subset of series that are affected by each most recent changepoint.
We measure the accuracy of the estimate of the number of most recent changepoints using the absolute error, $\lvert \hat{K} - K \rvert$, and call this the changepoint accuracy. 
We then measure the accuracy of the estimates of the subsets of series affected by each of the changepoints using the set coverage 
\[
  D_j = 1 - \frac{ \lvert \hat{I}_j \cap I_j  \rvert }{ \sqrt{ \lvert  \hat{I}_j \rvert \lvert I_j \rvert } }.
\]
Here $I_j$ is the true subset of series affected by the $j$ most recent changepoint and $\hat{I}_j$ is the estimated subset.
This measure satisfies 
$D_j \in [0,1]$, with $D_j=0$ indicating that the estimated subset overlaps exactly with the true subset, and $D_j =  1$ if the two subsets are disjoint.
More generally, smaller values of $D_j$ indicate a greater overlap. In the simulations presented for each panel we calculate the mean of  $D_1 , D_2 , \hdots , D_{\hat{K}}$. 
    
{\bf Case 1. Effect of $K$.}

For the first study we simulated data as described above for a range of values for $K$ from $K=1$ to $K=10$. Results are shown in Table \ref{tab:ss_mult}.

\begin{table}[h!]
  \begin{center}
    \begin{tabular}{c|cc|cccc|cccc}
      \hline\hline 
                          &     \multicolumn{2}{|c|}{IND} & \multicolumn{4}{|c|}{DCBS}  & \multicolumn{4}{|c}{MRC}  \\    
                                                                    \cline{2-11}
                    $k$     &  PD   &  LA   & PD   &  CA   &  LA   &  D   &  PD   &  CA  &   LA   &   D   \\ [0.5ex] 
      \hline 
1 & 0.73 & 1.46 & 0.86 & 2.98 & 0.05 & 0.09 & 0.98 & 0.10 & 0.06 & 0.01 \\ 
  2 & 0.76 & 1.43 & 0.82 & 2.55 & 0.04 & 0.14 & 0.97 & 0.04 & 0.04 & 0.03 \\ 
  3 & 0.77 & 1.39 & 0.70 & 2.64 & 0.13 & 0.24 & 0.95 & 0.05 & 0.03 & 0.05 \\ 
  4 & 0.77 & 1.37 & 0.67 & 2.47 & 0.18 & 0.28 & 0.94 & 0.03 & 0.05 & 0.06 \\ 
  5 & 0.78 & 1.38 & 0.58 & 2.09 & 0.24 & 0.35 & 0.93 & 0.03 & 0.04 & 0.07 \\ 
  10 & 0.78 & 1.41 & 0.29 & 1.77 & 0.76 & 0.62 & 0.89 & 0.10 & 0.19 & 0.10 \\ 
      \hline\hline
    \end{tabular}
  \end{center}
  \caption{For all of the methods and differing values of $K$ we repeated each experiment 100 times and recorded the proportion of true changes we detected (PD), the accuracy in detecting the number of
  distinct most recent changes (CA), the accuracy of the estimated location of these changes (LA) and the set coverage (D). These values are averaged over the 100 replications.}
  \label{tab:ss_mult}
\end{table}    


It is clear from Table \ref{tab:ss_mult} that our MRC method outperforms both IND and DCBS across the criteria we consider. The ability to synthesise information across time series means that MRC is able 
to more accurately detect changes and locate where they occur than analysing each time series independently. Not surprisingly we see that the advantage of using MRC over IND decreases as $K$ increases. We
also see that DCBS is more accurate than IND for small values of $K$, and is consistently more accurate in estimating the position of detected changepoints, but appears less powerful at detecting the most recent changes
as $K$ increases.

{\bf Case 2. Effect of size of change at final changepoint.}
\label{sec:case2}

Next we look at how the performance of each method is affected by the size of the mean change at the most recent changepoint, $\epsilon$. We fix the number of most recent changepoints as $K=5$, meaning that there are 20 series affected by each different changepoint. We vary the value of $\epsilon$ from $\epsilon = 0.2$ to $\epsilon = 1.6$. 
Results are shown in Table \ref{tab:ss_eps}. 

We again see MRC giving consistently stronger performance for all values of $\epsilon$. The advantage of MRC over IND is largest for moderate values of $\epsilon$. For 
small values of $\epsilon$ the information about changes in each time series is small, and thus the benefit of merging information across time series is limited. For larger values of $\epsilon$ it is relatively
easy to detect changes from an individual time series, and hence the benefit of using MRC over IND is mainly seen in its ability to more accurately locate the position. Surprisingly DCBS does not improve
as much as the other methods as we increase $\epsilon$. The DCBS method was not specifically designed to 
detect most recent changes, and it appears not to be as accurate at identifying which time series change at each changepoint, which then impacts its accuracy at detecting which changes are most 
recent for a given time series. 

\begin{table}[t!]
  \begin{center}
    \begin{tabular}{c|cc|cccc|cccc}
      \hline\hline 
                          &     \multicolumn{2}{|c|}{IND} & \multicolumn{4}{|c|}{DCBS}  & \multicolumn{4}{|c}{MRC} \\    
                                                                    \cline{2-11}
                    $\epsilon$     &  PD   &  LA  &   PD  &  CA &  LA  &  D   & PD  &  CA &  LA  &  D    \\ [0.5ex] 
      \hline 
0.2 & 0.11 & 1.49 & 0.09 & 3.49 & 0.75 & 0.66 & 0.11 & 2.06 & 0.47 & 0.64 \\ 
  0.4 & 0.22 & 1.75 & 0.22 & 3.43 & 1.01 & 0.55 & 0.36 & 1.27 & 0.88 & 0.42 \\ 
  0.6 & 0.46 & 1.76 & 0.37 & 3.00 & 0.64 & 0.51 & 0.76 & 0.30 & 0.29 & 0.20 \\ 
  0.8 & 0.65 & 1.57 & 0.48 & 2.51 & 0.35 & 0.45 & 0.89 & 0.09 & 0.12 & 0.10 \\ 
  1 & 0.78 & 1.38 & 0.58 & 2.09 & 0.24 & 0.35 & 0.93 & 0.03 & 0.04 & 0.07 \\ 
  1.2 & 0.86 & 1.19 & 0.62 & 1.63 & 0.16 & 0.31 & 0.95 & 0.04 & 0.02 & 0.05 \\ 
  1.4 & 0.91 & 1.01 & 0.65 & 1.44 & 0.13 & 0.26 & 0.95 & 0.04 & 0.00 & 0.05 \\ 
  1.6 & 0.93 & 0.85 & 0.66 & 1.47 & 0.10 & 0.25 & 0.96 & 0.04 & 0.00 & 0.04 \\ 
      \hline\hline
    \end{tabular}
  \end{center}
  \caption{For all of the methods with a fixed value of $K=5$ and differing values of $\epsilon$ we repeated each experiment 100 times and recorded the proportion of true changes we detected (PD), 
  the accuracy in detecting the number of
  distinct most recent changes (CA), the accuracy of  the estimated location of these changes (LA) and the set coverage (D). These values are averaged over the 100 replications.}
  \label{tab:ss_eps}
\end{table}    




{\bf Case 3. Dependent observations.}

One of the key assumptions we made when modelling the most recent change process was the independence of observations, both within and between segments. 
This greatly simplifies the modelling and especially the inference procedure. However, in many real time series applications observations are not independent and display serial autocorrelation.       

To assess the robustness of the MRC procedure we simulated an MRC process with a piecewise constant mean function as before, but instead of adding IID normally distributed `noise' we simulated an
$AR(1)$ noise process, $Z_t$, with standard normal errors $e_t$   
\begin{align*}
Z_t = \phi Z_{t-1} + e_t.
\end{align*}
This process was simulated for a range of values of $\phi$ which represented mild to moderate autocorrelation. The number of most recent changepoints was fixed at $K=5$ and we set $\epsilon = 1$.  Results are shown in Table \ref{tab:ss_dep}. 

\begin{table}[h]
  \begin{center}
    \begin{tabular}{c|cc|cccc|cccc}
      \hline\hline 
                          &     \multicolumn{2}{|c|}{IND} & \multicolumn{4}{|c|}{DCBS}  & \multicolumn{4}{|c}{MRC}   \\    
                                                                    \cline{2-11}
                    $\phi$     &  PD   &  LA   & PD   &  CA   &  LA   &  D   &  PD   &  CA  &   LA   &   D    \\ [0.5ex] 
      \hline 
-0.4 & 0.90 & 1.16 & 0.62 & 1.92 & 0.16 & 0.29 & 0.98 & 0.02 & 0.05 & 0.02 \\ 
  -0.3 & 0.87 & 1.22 & 0.62 & 1.97 & 0.18 & 0.30 & 0.98 & 0.02 & 0.03 & 0.02 \\ 
  -0.2 & 0.85 & 1.29 & 0.60 & 2.19 & 0.23 & 0.33 & 0.97 & 0.02 & 0.05 & 0.03 \\ 
  -0.1 & 0.83 & 1.33 & 0.58 & 2.16 & 0.30 & 0.35 & 0.96 & 0.02 & 0.04 & 0.04 \\ 
  0 & 0.78 & 1.38 & 0.58 & 2.09 & 0.24 & 0.35 & 0.93 & 0.03 & 0.04 & 0.07 \\ 
  0.1 & 0.73 & 1.47 & 0.55 & 2.17 & 0.29 & 0.40 & 0.89 & 0.03 & 0.07 & 0.11 \\ 
  0.2 & 0.64 & 1.56 & 0.43 & 2.66 & 0.31 & 0.49 & 0.75 & 0.74 & 0.18 & 0.21 \\ 
  0.3 & 0.51 & 1.59 & 0.12 & 3.05 & 0.66 & 0.71 & 0.47 & 1.71 & 0.35 & 0.38 \\ 
  0.4 & 0.36 & 1.72 & 0.04 & 2.70 & 1.59 & 0.80 & 0.21 & 1.79 & 0.82 & 0.55 \\ 
      \hline\hline
    \end{tabular}
  \end{center}
  \caption{For all of the methods and differing values of $\phi$ we repeated each experiment 100 times and recorded the proportion of true changes we detected (PD), the accuracy in detecting the number of
  distinct most recent changes (CA), the accuracy of estimated location of these changes (LA) and the set coverage (D). These values are averaged over the 100 replications. Fixed values for $K=5$ and $\epsilon = 1.0$ were used.}
  \label{tab:ss_dep}
\end{table}    

As $\phi$ increases the dependence between observations increases and the measures for all methods we consider decrease. The impact on both MRC and DCBS is larger
than the impact on IND, with IND correctly detecting more recent changepoints for $\phi=0.4$. Both MRC and DCBS still give more accurate estimates of the position of the changes that they do detect, and MRC is again
more accurate than DCBS for all cases we consider.


We then simulated an $MA(1)$ noise process, $Z_t$, with standard normal errors $e_t$   
\begin{align*}
Z_t =  e_t + \phi e_{t-1}.
\end{align*}
This process was simulated for a range of values of $\phi$. The number of most recent changepoints was fixed at $K=5$ and we set $\epsilon = 1$.  Results are shown in Table \ref{tab:ss_depMA}. We see similar patterns
to those observed for the $AR(1)$ model. Increased autocorrelation reduces the accuracy of all methods. For the largest values of $\phi$ we tried, IND performs slightly better than MRC in terms of the proportion
detected, but MRC and DCBS still give more accurate estimates of the location of the changes they do detect. For all cases MRC outperforms DCBS.

The fact that increasing the level of autocorrelation in the residuals, for both the $AR(1)$ and $MA(1)$ models, reduces the accuracy of all methods is not surprising. As we increase the autocorrelation there will be
less information in the data about the position of changes. Furthermore, all methods were implemented with a choice of penalty that assumes IID residuals. When there is positive autocorrelation this can lead to
such methods detecting a larger number of spurious changepoints. Increasing the penalty or threshold that defines when we add a change can combat this effect \cite[see][]{Lavielle2000},
and it may be that slightly better performance of all methods can be obtained by adapting the penalty or threshold in line with the level of autocorrelation in the residuals.

\begin{table}[h!]
  \begin{center}
    \begin{tabular}{c|cc|cccc|cccc}
      \hline\hline 
                          &     \multicolumn{2}{|c|}{IND} & \multicolumn{4}{|c|}{DCBS}  & \multicolumn{4}{|c}{MRC}   \\    
                                                                    \cline{2-11}
                    $\phi$     &  PD   &  LA   & PD   &  CA   &  LA   &  D   &  PD   &  CA  &   LA   &   D    \\ [0.5ex] 
      \hline 
-0.4 & 0.92 & 1.11 & 0.63 & 1.87 & 0.25 & 0.27 & 0.99 & 0.04 & 0.03 & 0.01 \\ 
  -0.3 & 0.90 & 1.17 & 0.63 & 2.00 & 0.22 & 0.29 & 0.98 & 0.04 & 0.03 & 0.02 \\ 
  -0.2 & 0.87 & 1.25 & 0.62 & 2.04 & 0.21 & 0.31 & 0.97 & 0.04 & 0.03 & 0.02 \\ 
  -0.1 & 0.83 & 1.31 & 0.61 & 2.21 & 0.24 & 0.33 & 0.96 & 0.04 & 0.03 & 0.04 \\ 
  0 & 0.78 & 1.38 & 0.58 & 2.09 & 0.24 & 0.35 & 0.93 & 0.03 & 0.04 & 0.07 \\ 
  0.1 & 0.73 & 1.45 & 0.56 & 2.23 & 0.27 & 0.39 & 0.89 & 0.04 & 0.06 & 0.11 \\ 
  0.2 & 0.67 & 1.50 & 0.50 & 2.46 & 0.31 & 0.45 & 0.81 & 0.37 & 0.09 & 0.17 \\ 
  0.3 & 0.60 & 1.57 & 0.27 & 2.78 & 0.50 & 0.61 & 0.65 & 1.32 & 0.23 & 0.28 \\ 
  0.4 & 0.51 & 1.65 & 0.09 & 2.70 & 0.80 & 0.73 & 0.45 & 1.78 & 0.40 & 0.39 \\ 
      \hline\hline
    \end{tabular}
  \end{center}
  \caption{For all of the methods and differing values of $\phi$ we repeated each experiment 100 times and recorded the proportion of true changes we detected (PD), the accuracy in detecting the number of
  distinct most recent changes (CA), the accuracy of the estimated location of these changes (LA) and the set coverage (D).
   These values are averaged over the 100 replications. Fixed values for $K=5$ and $\epsilon = 1.0$ were used.}
  \label{tab:ss_depMA}
\end{table}

{\bf Case 4. Accuracy of prediction.}

Finally, we consider how each method performs if the aim is to predict $Y_{i,n+1},\ldots,Y_{i,n+5}$ for each time series. Each method gives an estimate for the most recent changepoint for each time series. Conditional on this
estimate we can estimate the mean in the final segment. This estimated mean is our prediction for the next value(s). We use the same data as in Case 1 but leave out 5 time points at the end of the data. We then predict the last 5 points using the most recent changepoints found by each method and 
measure the Mean Squared Error (MSE) between the truth and our predictions.  Results are shown in Table \ref{tab:ss_pred}. 
 
\begin{table}[h!]
  \begin{center}
    \begin{tabular}{c|c|c|c|c|c|c}
      \hline\hline 
                      $k$    &  IND  &  \multicolumn{1}{|c|}{AGG}  &   \multicolumn{1}{|c|}{MV}  &  \multicolumn{1}{|c|}{ECP}   & \multicolumn{1}{|c|}{DCBS} & \multicolumn{1}{|c}{MRC}  \\    
                                                                    \cline{2-7}
      \hline 
1 & 1.04 & 1.29 & 1.01 & 1.09 & 1.08 & 1.01 \\ 
  2 & 1.06 & 1.27 & 1.05 & 1.09 & 1.08 & 1.03 \\ 
  3 & 1.04 & 1.25 & 1.06 & 1.08 & 1.07 & 1.02 \\ 
  4 & 1.04 & 1.34 & 1.08 & 1.09 & 1.08 & 1.02 \\ 
  5 & 1.04 & 1.23 & 1.08 & 1.09 & 1.07 & 1.02 \\ 
  10 & 1.04 & 1.29 & 1.12 & 1.09 & 1.09 & 1.02 \\ 
      \hline\hline
    \end{tabular}
  \end{center}
  \caption{The average Mean Squared Error (MSE) for predictions of each method. The MSE was calculated for the difference between the truth and predicted values and averaged over 100 replications.}
  \label{tab:ss_pred}
\end{table}    

MRC gives the most accurate predictions for all values of $K$, and is the only method to consistently be more accurate than analysing each time series individually. 
The method which treats the $N$ time series as a multivariate time series, where the mean changes in all components at a change (MV), does well for $K=1$ and $K=2$, but loses accuracy for larger $K$. The method
that aggregates the time series, and then detects changes in the resulting uni-variate time series, does particularly poorly. This is because the aggregation step reduces the signal for a change, even when all changes
are in the same location, as the sign of the change in mean differs across time series.



\section{Applications}

We look at two different applications of our method using real data. These applications differ in their focus and the aim of the analysis. 
The first is that of the telecommunications event count data introduced in Section \ref{sec:intro}.
Our second application concerns the balance sheets of a large number of firms. In this latter case we look for changes in a parameter that measures the ratio between the cash holdings of a company and the net assets held on its balance sheet. The goal of this analysis is to explore why the cash holdings of many large firms have increased over time, and if there are any specific events which have caused this. 
By using our method we can identify the years when a change occurs and for each of these years, which firms change. This information helps us to tie in specific legal or economic changes to the years in which they happened and the types of industries that are affected.

\subsection{Telecommunications event data}
\label{sec:BT}

Our panel data consists of the number of events that occur each week over a 175 week period. Events are recorded for each of 10 geographical regions and 16 different event types. 
 Thus there are 160 possible series, of which 18 of these show no weekly events over the 175 weeks measured. So we are left with 142 series to analyse.

\begin{figure}[h!]
  \centering
  \includegraphics[scale=0.4]{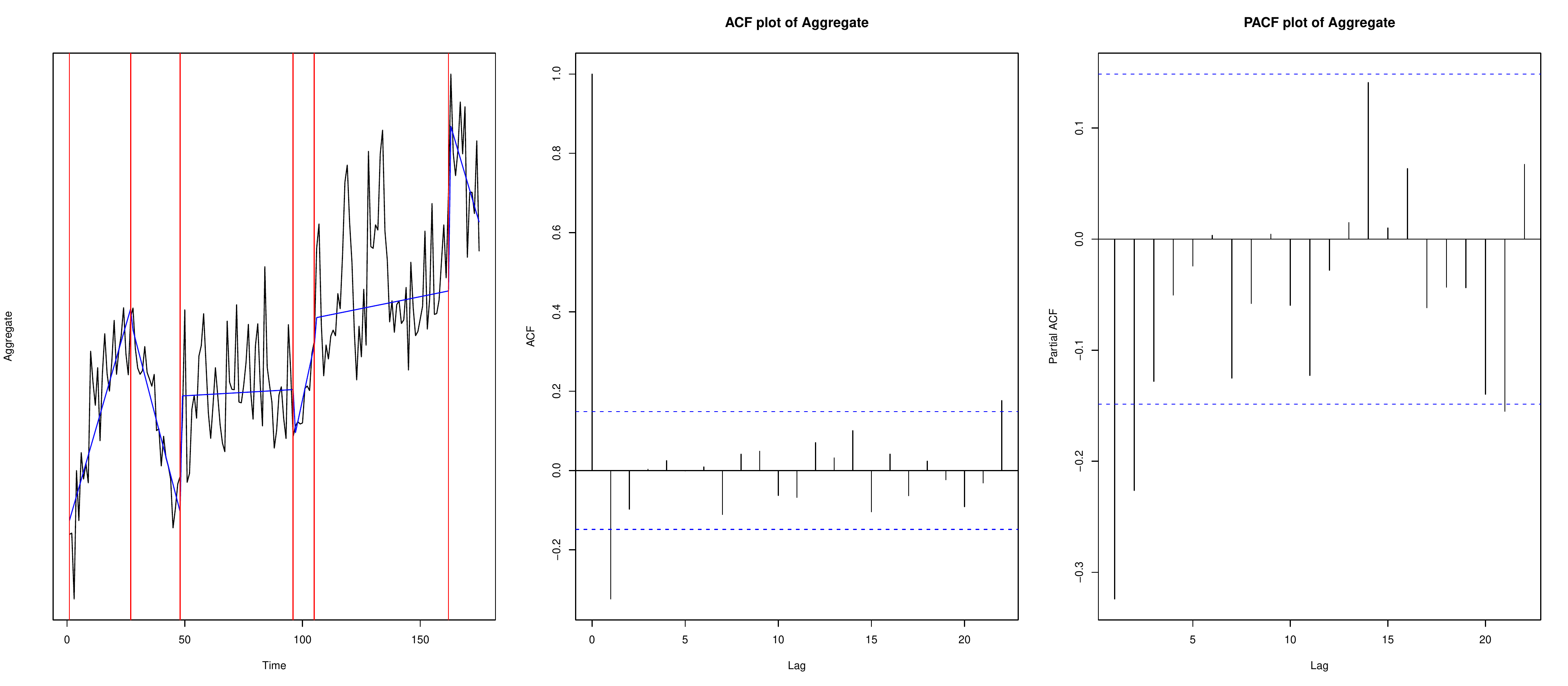}
  \caption{The aggregate series segmented into piece wise linear regressions and ACF, PACF plots of the first differences of the aggregate series.}
  \label{fig:agg_bt}
\end{figure}

We can get an overall time series for the number of events per week across the entire network if we aggregate all of these series together. 
This fully aggregated series is shown in Figure \ref{fig:agg_bt}. We can see that there are distinct changes in the slope of this series and it is segmented into piecewise linear regressions. 
 We can get some 
idea of the level of auto-correlation within the data by calculating the autocorrelation and partial-autocorrelation for this aggregated data after differencing the data to remove the trend component. Plots
of the resulting ACF and PACF plots are shown in Figure \ref{fig:agg_bt}, and they suggest the residuals could be modelled by an $MA(1)$ with negative autocorrelation -- a situation where we saw MRC performing well in the simulation
study.

As mentioned in the introduction, the main interest with this data is in making short-term predictions. To do this we use the method described in Section \ref{sec:mmrc} to find the number of most recent changes.
Our method is applied assuming the mean of the data within each segment is a linear function of time, and  using the robust cost function (\ref{eq:Robust}).
We estimate that there are five different most recent changepoints. This means that all of the 142 series can be separated into five groups depending upon which of the five most recent changepoint affects each series. 


\begin{figure}[h!]
  \centering
\includegraphics[width=.3\textwidth]{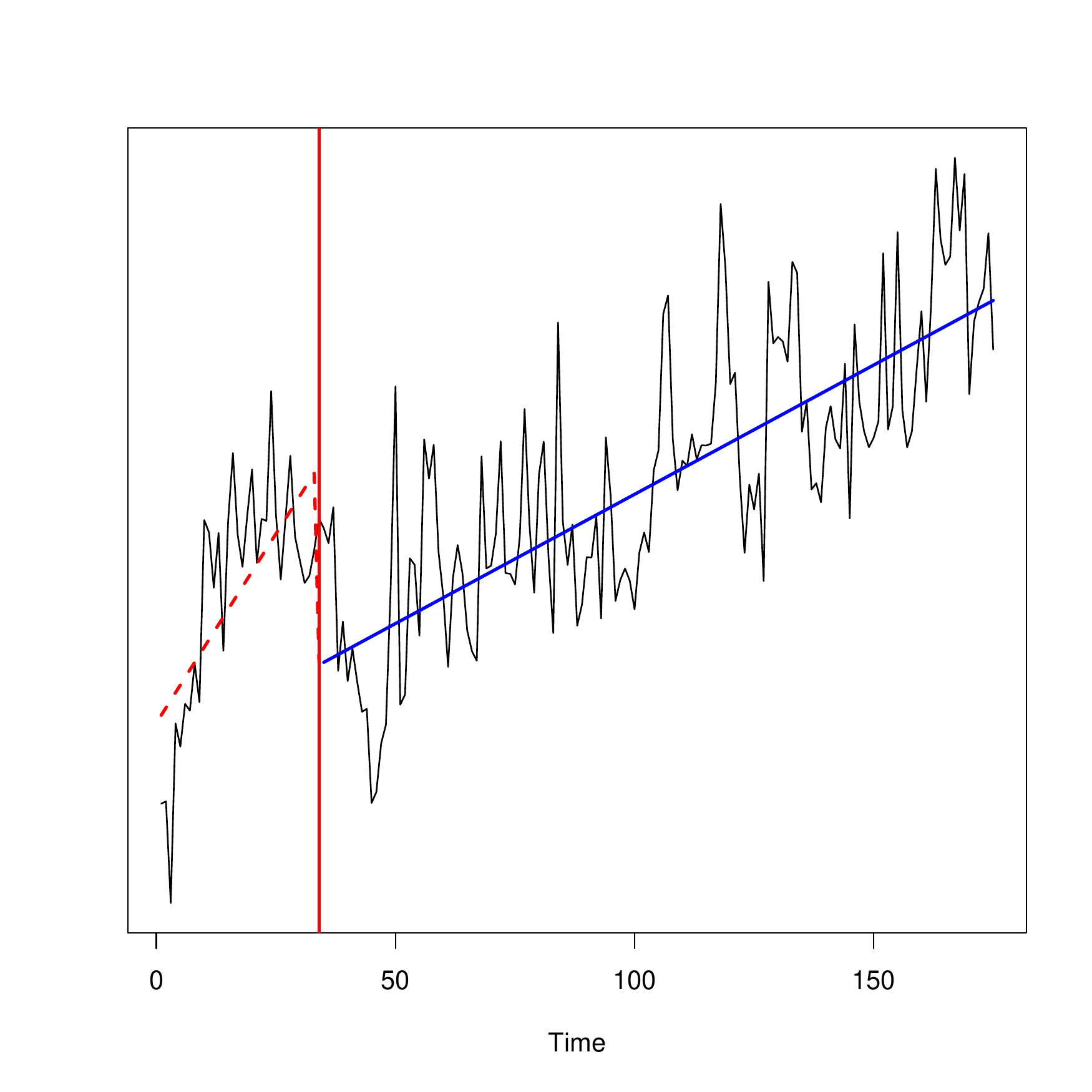}\quad
\includegraphics[width=.3\textwidth]{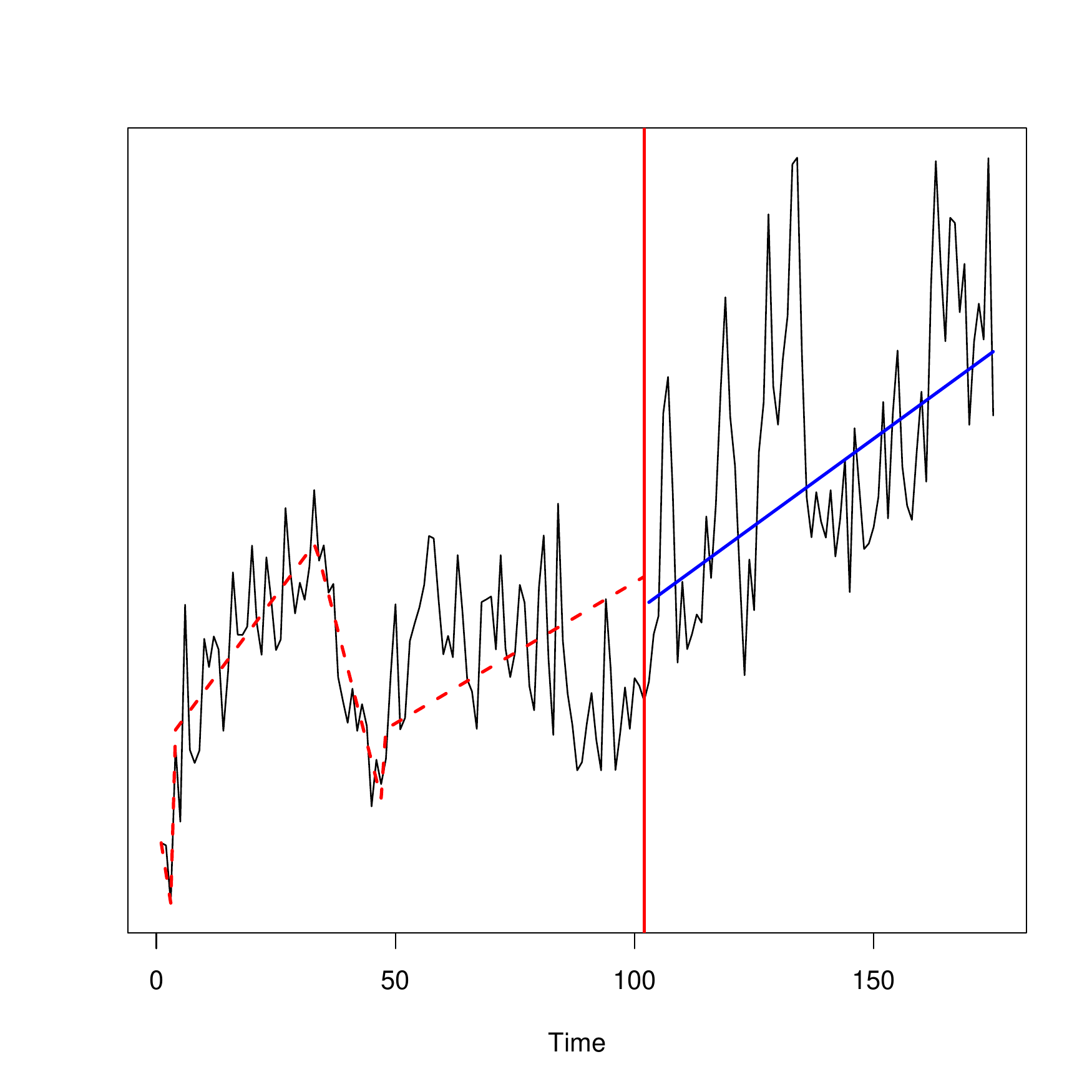}\quad
\includegraphics[width=.3\textwidth]{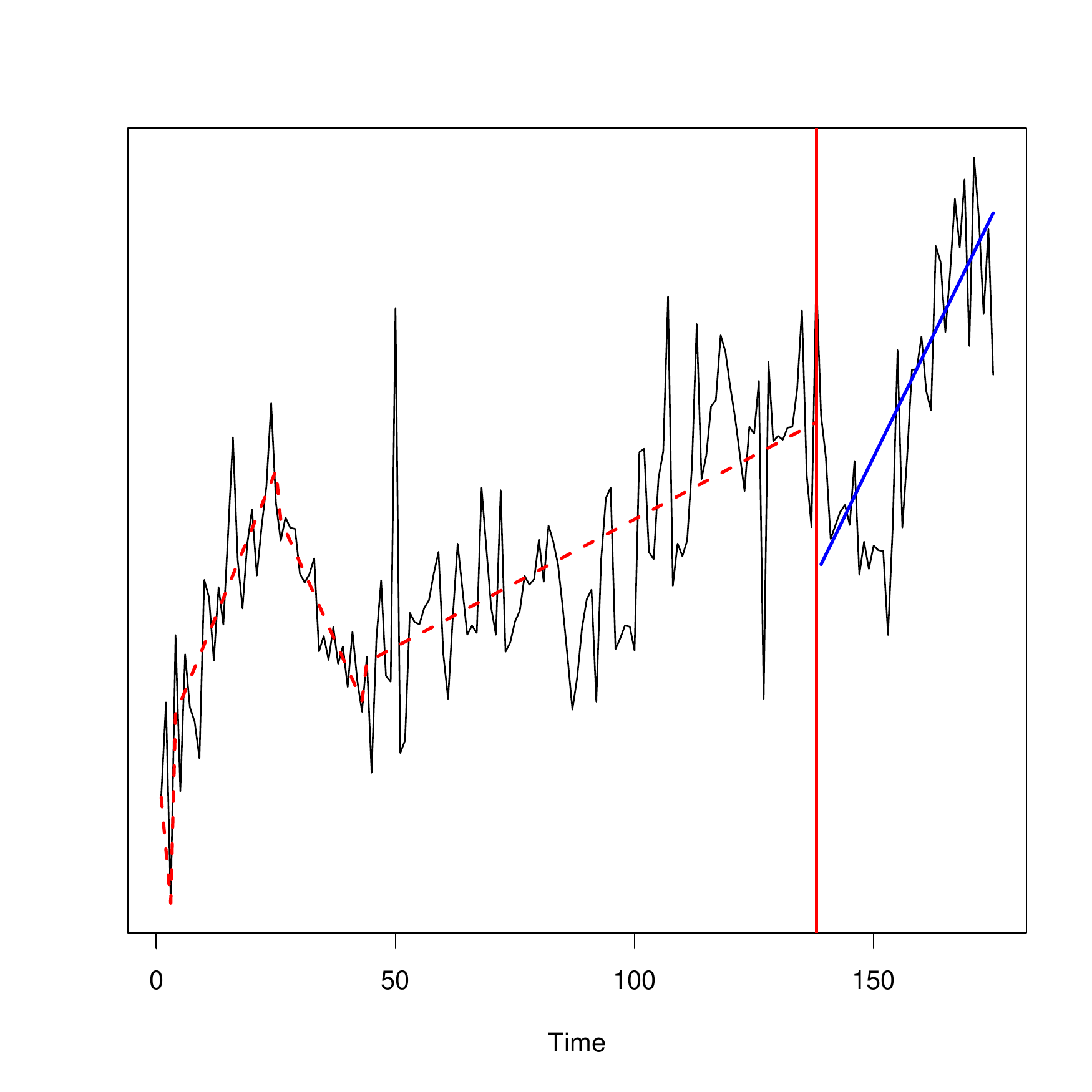}
\medskip
\includegraphics[width=.3\textwidth]{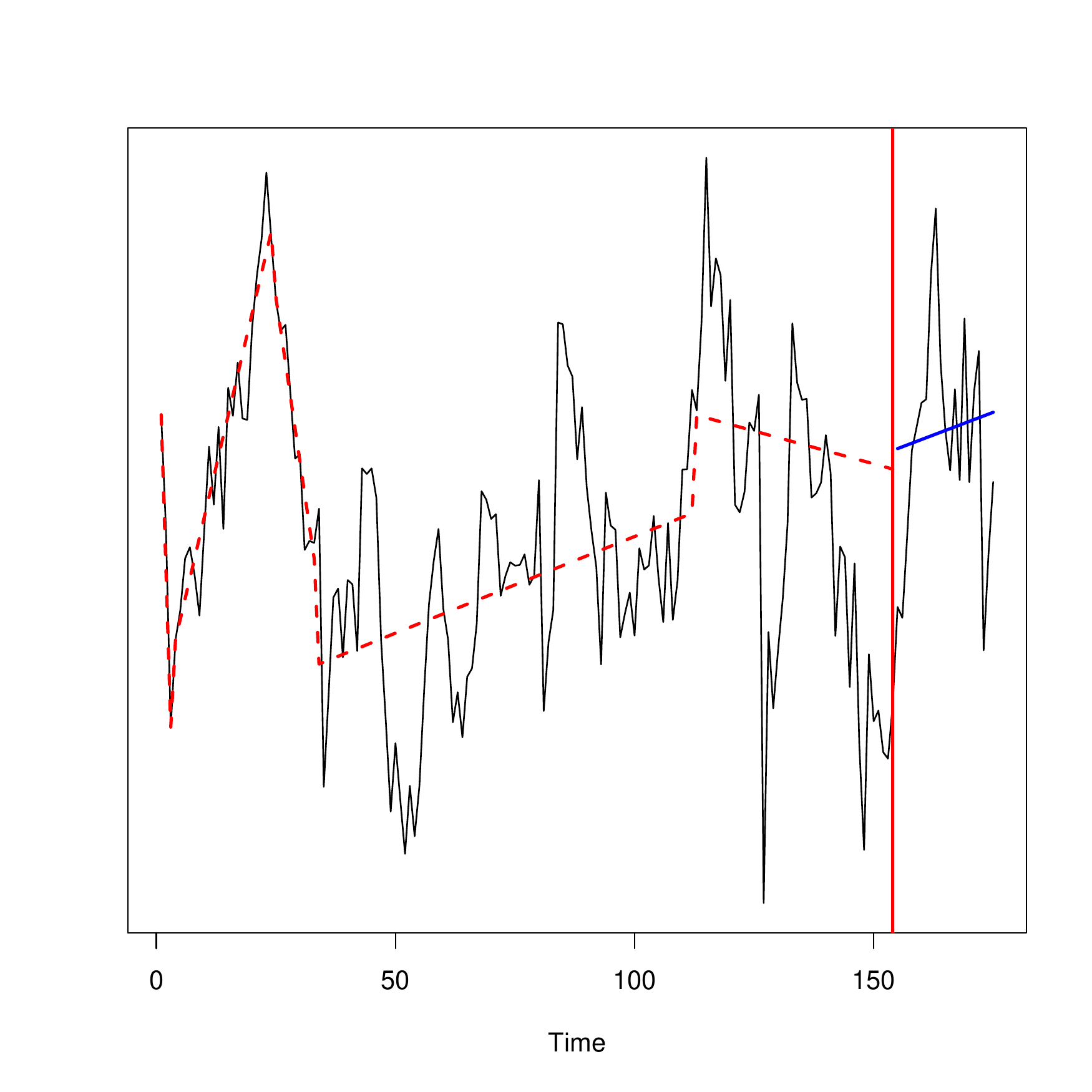}\quad
\includegraphics[width=.3\textwidth]{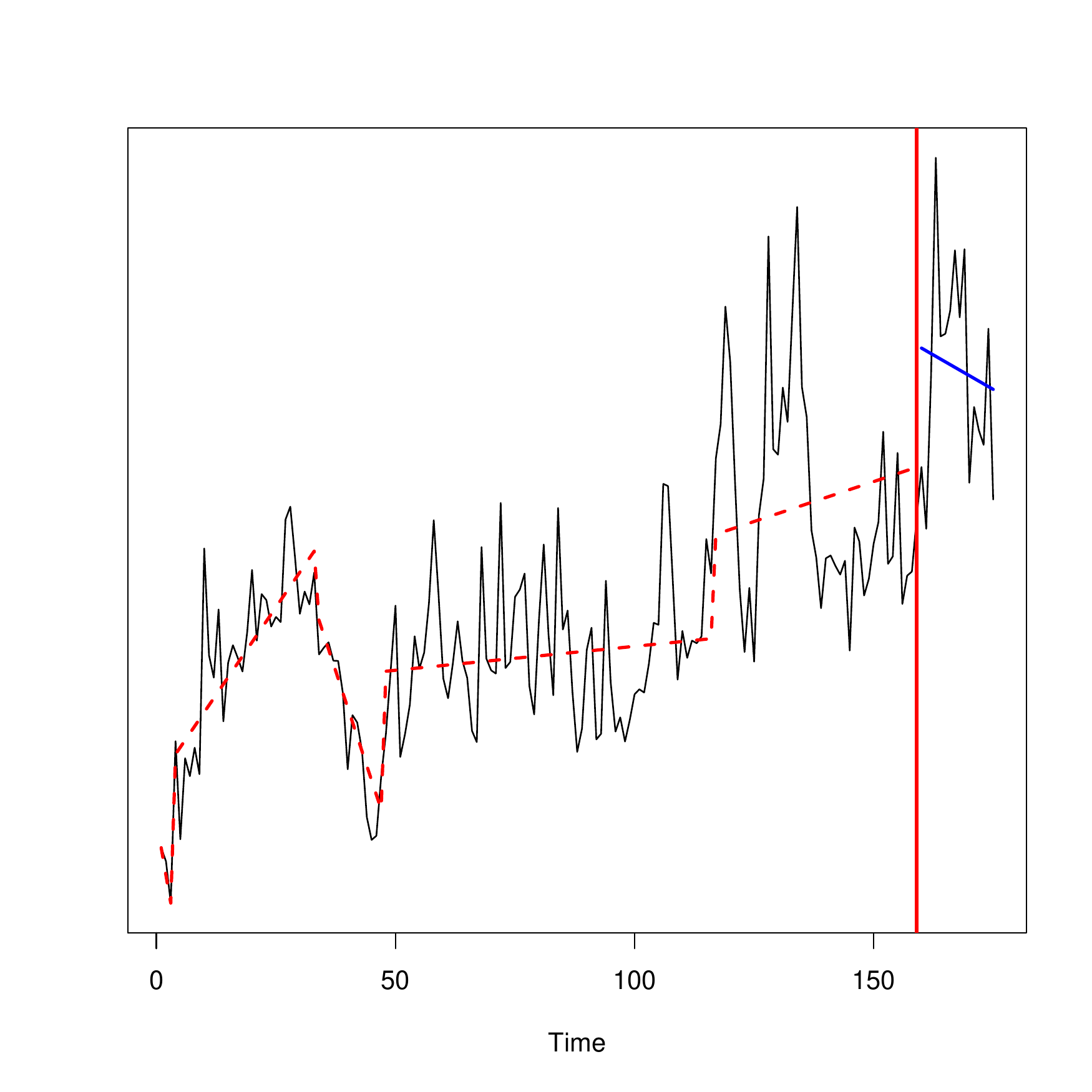}
  \caption{The aggregate series for each of the five groups of series. Their respective most recent changepoints are added, with the final segment shown in blue. The previous segmentation prior to the most recent change is shown as a red dashed line.}
  \label{fig:agg_mrcs}
\end{figure}

Figure \ref{fig:agg_mrcs} show the aggregate series for each of the five groups. The groups contain 26, 27, 28, 28 and 33 series from left to right respectively. 

All of the aggregated series show an increased trend initially until around the 35$^{th}$ week. This can be seen most prominently in the first series on the left, with a lower consistent gradient after this change. The second series shows that at around the 100$^{th}$ week the gradient of the trend increases slightly.
In the third series at around the 140$^{th}$ week the gradient of the final segment increases markedly. 
The fourth and fifth series both show a most recent change which is close to the end of the series, at around the 160$^{th}$ week, with a marked decrease in trend for the fifth series.

We can see several characteristics of the fully aggregated series in Figure \ref{fig:agg_bt} ``stripped'' almost into their component parts. The fourth series is somewhat of an anomaly as it is highly variable, upon further inspection this set of series was made up of individual series which all contained a small number of events per week and were quite variable.


When we have found the most recent changepoints, the parameters of the resulting regression line in the last segment can be estimated. 
These estimates can then be used to predict succeeding time points.
We analyse the data up to four data points (weeks) from the end of the data and then use the predictions obtained from the regression model to evaluate 
the Mean Square Error (MSE) of the prediction for the last four weeks. 

We compare predictions using the estimated most recent changepoints from MRC with predictions where we segment each time series separately.
The MSE for the predictions in the latter case is 
43442 while for our algorithm (with $K=5$) it is 41779. This is an improvement of 3.8\% in the MSE of the prediction compared to analysing each time series individually.

\subsection{Corporate finance data}
\label{sec:corp_fin}

We now apply our method to a panel data set from the field of Corporate finance. This data set comprises the annual value of a range of different financial indicators for a number of firms.
These include, for example, the value of a firm's assets or whether the firm pays dividends or not.
This particular data set is known as an unbalanced panel as the observations for each firm do not all begin or end in the same year. We can view this as a longitudinal data problem where the cohort are the firms that are tracked over time. As is common in these problems there is a large  (cohort) number of firms, 7039 in this example, but these are observed over a much smaller time frame. In this case there are a maximum of 53 observations per firm (annually from 1962 - 2015).      

An intriguing phenomenon in corporate finance is the fact that U.S. firms hold considerably more cash nowadays compared with a few decades previously.  
Specifically, cash as a proportion of total assets held by U.S. firms has more than doubled in the past three decades.  
The evolution of corporate cash holdings has received a lot of attention from academic researchers, policy makers, and practitioners.  
Numerous explanations for this have been offered in the literature, including increased cash flow volatility \citep{bates2009,bates2011}, 
competition \citep{brown2011}, changes in production technology \citep{gao2013} and changes in the cost of carry \citep{schmalz2015}.  

\cite{schmalz2015} argue that changes over time in the cost of carry, that is the net cost of financing one dollar of liquid assets, explains the evolution of corporate cash holdings \cite[see also][]{graham2015}. 
They measure the cost of carry as the spread between the risk-free Treasury-bill rate and the return on the portfolio of liquid assets for the corporate sector.  However a limitation of existing studies is that they split their data along the time domain into distinct `regimes' by eyeballing the data.  Such an approach is highly subjective and increases the opportunity for data snooping.  It would be preferable to introduce a formal procedure for detecting any distinct regimes.

We therefore re-examine the ability of the cost of carry to capture variation in corporate cash by formally modelling the breakpoint process using our changepoint methodology. 
Our analysis follows \cite{schmalz2015} and therefore uses the same dataset \cite[see][for a detailed description of the dataset]{schmalz2015}.  
We control for a number of variables that may affect cash holdings of a firm, such as capital expenditure, spending on R\&D and the amount of leverage it has amongst others. 
Specifically we consider a fixed effects linear model where the response variable, $y_{it}$, represents the cash to net asset ratio of firm $i$ in year $t$ is regressed against $12$ covariates,
\begin{align}
\label{eq:felm}
y_{it} = \alpha_i +  \beta_1 X_{1it} + \beta_2 X_{2it} + \hdots + \beta_{12} X_{12it} + \epsilon_{it}.
\end{align}
These covariates are described in Table \ref{tab:description}.
The $\beta_j$s are pooled estimates of the effect of the covariates measured over all 7039 firms and the years in which they are observed. 
%
Each fixed effect term, $\alpha_i$, captures a firm-specific characteristic in terms of a firm specific intercept. These fixed effects can be interpreted as the difference 
between the predicted cash to net assets ratio and the true value observed. As such, the fixed effects are able to capture differences caused by external changes which cannot be explained by the covariates
in the model.  



For a specific firm the fixed effects term may change due to a number of factors such as a CEO change, a merger or takeover by another firm or some scandal such as a product 
recall which requires large amounts of cash to be spent. However, we are more interested in the times at which the fixed effect parameter changes in a significant number of firms at the same  time. 
The causes of these changes would be due to wider economic events such as changes in policy, technological innovation,  or regulatory changes such as the Sarbanes-Oxley Act of 2002.  

Having estimated the $\beta_j$s via maximum likelihood estimate, we can rewrite \eqref{eq:felm} as a change in mean model
\begin{align}
  y_{it} - \left(  \hat{\beta}_1 X_{1it} + \hat{\beta}_2 X_{2it} + \hdots + \hat{\beta}_{12} X_{12it} \right) = \alpha_{it} + \epsilon_{it},
\end{align}
where the $\hat{\beta}_i$s are the parameter estimates.

Our MRC method can be applied to this problem and aims to find the year(s) in which the most recent changepoint(s) occur and the subsets of firms that are affected. 
We now follow the method of Section \ref{sec:mmrc} to find the optimal number of most-recent changepoints and the sets of firms that are affected by them.

\subsubsection{The Estimated Changepoints}

We find three most-recent changepoints. These are located in years 1979, 1996 and 2007. The largest subset, approximately 70\%, of firms have their most recent change at 1979.  
This date corresponds to a change in the Federal Reserve's operating procedures;
specifically it marks the beginning of the `monetarist policy experiment', and is identified as a breakpoint in \cite{corpfin_strucbreak_79} who use a historical time series of excess returns that
are subject to breaks, to forecast the equity premium out-of-sample.  

Another benefit of our methodology is the ability to observe which firms are undergoing a change and which are not.  
This is of real interest in economic and finance applications because it may be able to provide information to help identify the underlying cause of the structural break.  
For instance if the change is experienced predominantly by firms in one industry it could be indicative of an industry-specific shock or regulation change.  Conversely,  
if the change occurs across all firms this might suggest an economy-wide change in policy.  In this application the affected firms are roughly equally distributed across each of the broader
industry classes strengthening the case for the cause being the change in the Federal Reserve's macroeconomic policy.  This policy change led to a decrease in the fixed effects part of the model 
in the majority of the firms that were affected by the change and thus a decrease in their cash holdings.

\begin{figure}[h]
  \centering
  \includegraphics[scale=0.4]{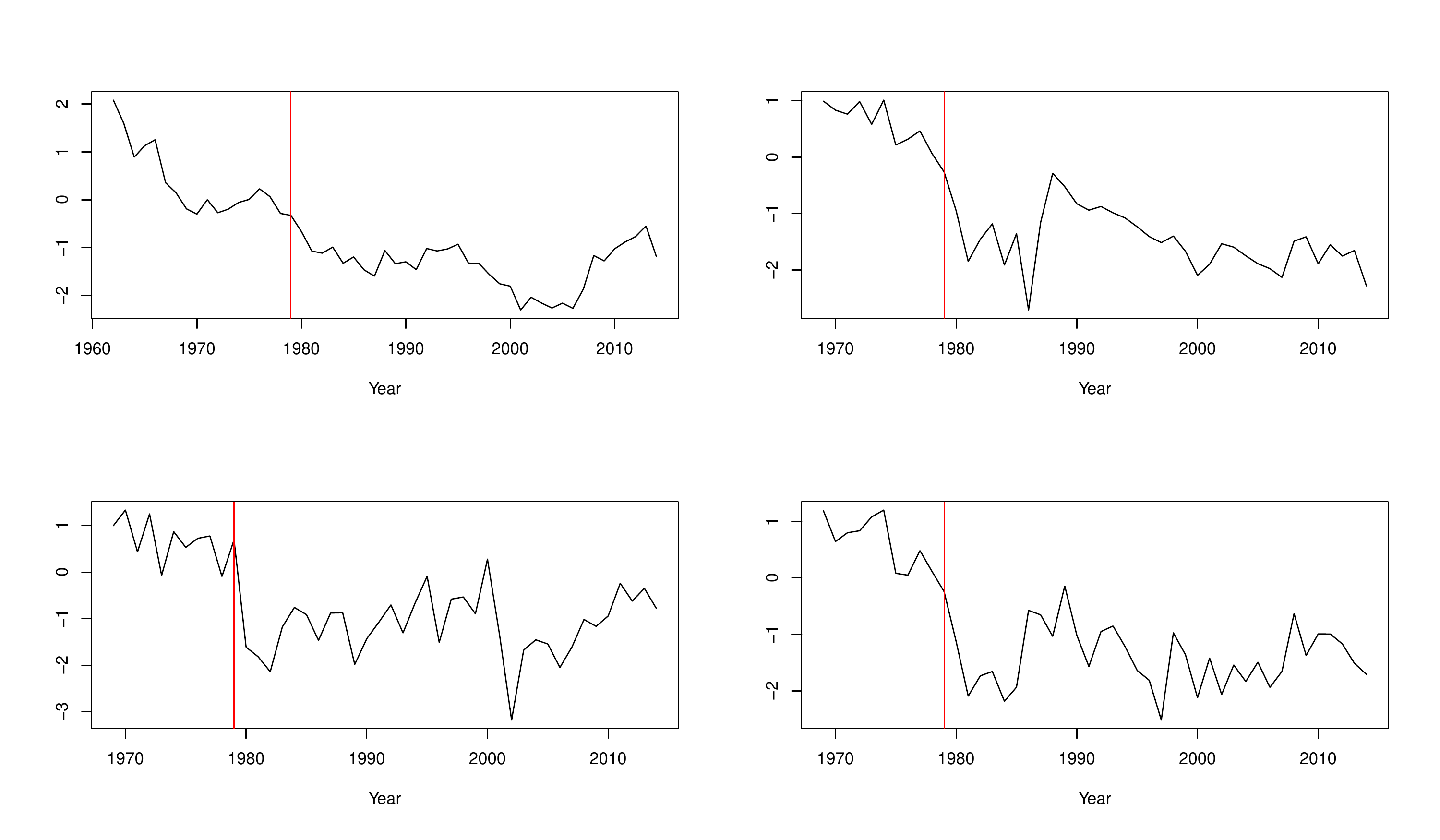}
  \caption{Some of the affected firms plots of their fixed effects showing a change in 1979.}
  \label{fig:mrc_1979}
\end{figure}

The changes at 1996 and 2007 each affect around 15\% of the firms.
The change in 1996 affected mostly Utilities firms and the one in 2007 affected both the Trade and Services sectors. 
The 1996 changepoint is also found in related work on structural breaks \citep{corpfin_strucbreak_79} and can be attributed to the late 1990's retail bull market in which net assets markedly increased in value.  
The Telecommunications Act of 1996 which deregulated the U.S. broadcasting and telecommunications markets could explain why the Utilities sector experienced a large shock.  
The deregulation paved the way for many utilities companies to enter the broadcasting and telecommunications market.  
The change in 2007 corresponds to the recent financial crisis and the large fluctuations in the value of assets held by many firms in that period.


\begin{table}[ht]
\centering
\begin{tabular}{c|c}
 \hline\hline 
  Covariate   &  Description       \\ 
  \hline
$X_{1it}$      &  T-Bill (the rate of return on a 90 day treasury bill)     \\
$X_{2it}$      &  Cost of carry       \\
$X_{3it}$      &  Log of real assets       \\
$X_{4it}$      &  Industry sigma (a measure of the volatility in each sector)        \\
$X_{5it}$      &  Cash flow to assets ratio    \\
$X_{6it}$      &  Net working capital to assets ratio      \\
$X_{7it}$      &  R\&D/Sales   \\
$X_{8it}$      &  Dividend dummy        \\
$X_{9it}$      &  Market to book ratio     \\
$X_{10it}$     &  Capital expenditure             \\
$X_{11it}$     &  Leverage          \\ 
$X_{12it}$     &  Acquistion activity             \\
   \hline\hline
\end{tabular}
  \caption{A description of the 12 covariates in the model.}
  \label{tab:description}
\end{table}

\newpage 
\section{Discussion}
In this paper we have developed novel methodology to detect changepoints in panel data. The specific changepoints we aim to detect are the most recent changes 
that affect different and disjoint subsets of the series that make up the panel. 
We focus on detecting the most recent changes as this can be useful in forecasting, as shown in Section \ref{sec:BT}. 
We are also able to identify which series are affected by different changes which leads to a greater understanding of why and how the changes have occurred. 

In our analysis of the two real data sets, we used cost functions for segmenting each individual time series that are based on assuming no temporal dependence in the residuals. This can be
justified theoretically by results that show, for example, that detecting changes in mean using a least squares criteria is robust to the presence of temporal dependence in the residuals
\cite[]{Lavielle2000}. We showed empirically that our method can still detect the most recent changes even in the presence of AR(1) structure. Furthermore, our general approach can easily be extended
to allow for modelling of the error structure of the residuals, by using cost functions for the data within each segment that are based on models which allow for autocorrelation.

Our method also ignores any dependence across time series, either in the form of cross-correlation in the residuals or of similar changes at common changepoints. 
Whilst the former is an active area of research within the non-stationary time series community \cite[see for example][]{Ombao05,Park14} this is an open 
and intriguing area of future research for the changepoint community. The consquence of ignoring such (time-varying) structure might be that we infer some spurious changes 
to fit unusual patterns in the residuals that are seen in multiple time series. It is not clear how to develop a method that accounts for the latter, but such a method could
have greater power at detecting changes than our MRC procedure.

{\bf Acknowledgements}
This research was supported by EPSRC grants EP/K014463/1 and EP/N031938/1. Bardwell gratefully acknowledges funding from EPSRC and BT via the STOR-i Centre for Doctoral Training.

\bibliography{bibliog}

\begin{thebibliography}{}

\bibitem[Azar et~al., 2016]{schmalz2015}
Azar, J., Kagy, J.-F., and Schmalz, M.~C. (2016).
\newblock Can changes in the cost of carry explain the dynamics of corporate
  cash holdings.
\newblock {\em Review of Financial Studies}, 29:2194–2240.

\bibitem[Bates et~al., 2017]{bates2011}
Bates, T.~W., Chang, C.~H., and Chi, J.~D. (2017).
\newblock Why has the value of cash increased over time?
\newblock {\em Journal of Financial and Quantitative Analysis (to appear)}.

\bibitem[Bates et~al., 2009]{bates2009}
Bates, T.~W., Kahle, K.~M., and Stulz, R.~M. (2009).
\newblock Why do {US} firms hold so much more cash than they used to?
\newblock {\em The Journal of Finance}, 64(5):1985--2021.

\bibitem[Brown and Petersen, 2011]{brown2011}
Brown, J.~R. and Petersen, B.~C. (2011).
\newblock Cash holdings and {R\&D} smoothing.
\newblock {\em Journal of Corporate Finance}, 17(3):694--709.

\bibitem[Cao and Wu, 2015]{cao2015changepoint}
Cao, H. and Wu, W.~B. (2015).
\newblock Changepoint estimation: another look at multiple testing problems.
\newblock {\em Biometrika}, 102(4):974--980.

\bibitem[Cho, 2016]{DCBS}
Cho, H. (2016).
\newblock Change-point detection in panel data via double cusum statistic.
\newblock {\em Electronic Journal of Statistics}, 10(2):2000--2038.

\bibitem[Cho and Fryzlewicz, 2015]{RSSB:RSSB12079}
Cho, H. and Fryzlewicz, P. (2015).
\newblock Multiple-change-point detection for high dimensional time series via
  sparsified binary segmentation.
\newblock {\em Journal of the Royal Statistical Society: Series B (Statistical
  Methodology)}, 77(2):475--507.

\bibitem[Davis et~al., 2006]{davis2006structural}
Davis, R.~A., Lee, T. C.~M., and Rodriguez-Yam, G.~A. (2006).
\newblock Structural break estimation for nonstationary time series models.
\newblock {\em Journal of the American Statistical Association},
  101(473):223--239.

\bibitem[Fearnhead and Rigaill, 2017]{Fearnhead/Rigaill:2016}
Fearnhead, P. and Rigaill, G. (2017).
\newblock Changepoint detection in the presence of outliers.
\newblock {\em Journal of the American Statistical Association (Forthcoming)}.

\bibitem[Fryzlewicz, 2014]{Fryzlewicz2012}
Fryzlewicz, P. (2014).
\newblock Wild binary segmentation for multiple change-point detection.
\newblock {\em The Annals of Statistics}, 42(6):2243--2281.

\bibitem[Gao, 2017]{gao2013}
Gao, X. (2017).
\newblock Corporate cash hoarding: The role of just-in-time adoption.
\newblock {\em Management Science (to appear)}.

\bibitem[Graham and Leary, 2016]{graham2015}
Graham, J.~R. and Leary, M.~T. (2016).
\newblock The evolution of corporate cash.
\newblock {\em SSRN}.
\newblock \mbox{doi}:{10.2139/ssrn.2805505}.

\bibitem[Gr{\"u}nwald, 2007]{grunwald2007minimum}
Gr{\"u}nwald, P.~D. (2007).
\newblock {\em The minimum description length principle}.
\newblock MIT press.

\bibitem[Haynes et~al., 2017]{CROPS}
Haynes, K., Eckley, I.~A., and Fearnhead, P. (2017).
\newblock Computationally efficient changepoint detection for a range of
  penalties.
\newblock {\em Journal of Computational and Graphical Statistics}, 26:134--143.

\bibitem[Jandhyala et~al., 2013]{JTSA:JTSA12035}
Jandhyala, V., Fotopoulos, S., MacNeill, I., and Liu, P. (2013).
\newblock Inference for single and multiple change-points in time series.
\newblock {\em Journal of Time Series Analysis}, 34:423--446.

\bibitem[Killick et~al., 2012]{pelt}
Killick, R., Fearnhead, P., and Eckley, I.~A. (2012).
\newblock Optimal detection of changepoints with a linear computational cost.
\newblock {\em Journal of the American Statistical Association},
  107(500):1590--1598.

\bibitem[Kirch et~al., 2015]{doi:10.1080/01621459.2014.957545}
Kirch, C., Muhsal, B., and Ombao, H. (2015).
\newblock Detection of changes in multivariate time series with application to
  {EEG} data.
\newblock {\em Journal of the American Statistical Association},
  110(511):1197--1216.

\bibitem[Lavielle, 2005]{Lavielle20051501}
Lavielle, M. (2005).
\newblock Using penalized contrasts for the change-point problem.
\newblock {\em Signal Processing}, 85(8):1501 -- 1510.

\bibitem[Lavielle and Moulines, 2000]{Lavielle2000}
Lavielle, M. and Moulines, E. (2000).
\newblock Least-squares estimation of an unknown number of shifts in a time
  series.
\newblock {\em Journal of Time Series Analysis}, 21(1):33--59.

\bibitem[Lavielle and Teyssi{\`e}re, 2006]{lavielle}
Lavielle, M. and Teyssi{\`e}re, G. (2006).
\newblock Detection of multiple change-points in multivariate time series.
\newblock {\em Lithuanian Mathematical Journal}, 46(3):287--306.

\bibitem[Ma and Yau, 2016]{ma2016pairwise}
Ma, T.~F. and Yau, C.~Y. (2016).
\newblock A pairwise likelihood-based approach for changepoint detection in
  multivariate time series models.
\newblock {\em Biometrika}, 103(2):409--421.

\bibitem[{Maidstone} et~al., 2017]{FPOP}
{Maidstone}, R., {Hocking}, T., {Rigaill}, G., and {Fearnhead}, P. (2017).
\newblock On optimal multiple changepoint algorithms for large data.
\newblock {\em Statistics and Computing}, 27:519--533.

\bibitem[Matteson and James, 2014]{ecp}
Matteson, D.~S. and James, N.~A. (2014).
\newblock A nonparametric approach for multiple change point analysis of
  multivariate data.
\newblock {\em Journal of the American Statistical Association},
  109(505):334--345.

\bibitem[Ombao et~al., 2005]{Ombao05}
Ombao, H., Von~Sachs, R., and Guo, W. (2005).
\newblock Slex analysis of multivariate nonstationary time series.
\newblock {\em Journal of the American Statistical Association},
  100(470):519--531.

\bibitem[Park et~al., 2014]{Park14}
Park, T., Eckley, I.~A., and Ombao, H.~C. (2014).
\newblock Estimating time-evolving partial coherence between signals via
  multivariate locally stationary wavelet processes.
\newblock {\em IEEE Transactions on Signal Processing}, 62(20):5240--5250.

\bibitem[Pettenuzzo and Timmermann, 2011]{corpfin_strucbreak_79}
Pettenuzzo, D. and Timmermann, A. (2011).
\newblock Predictability of stock returns and asset allocation under structural
  breaks.
\newblock {\em Journal of Econometrics}, 164(1):60--78.

\bibitem[Preuss et~al., 2015]{doi:10.1080/01621459.2014.920613}
Preuss, P., Puchstein, R., and Dette, H. (2015).
\newblock Detection of multiple structural breaks in multivariate time series.
\newblock {\em Journal of the American Statistical Association},
  110(510):654--668.

\bibitem[Reese, 2006]{NET:NET20128}
Reese, J. (2006).
\newblock Solution methods for the p-median problem: An annotated bibliography.
\newblock {\em Networks}, 48(3):125--142.

\bibitem[Teitz and Bart, 1968]{teitz1968heuristic}
Teitz, M.~B. and Bart, P. (1968).
\newblock Heuristic methods for estimating the generalized vertex median of a
  weighted graph.
\newblock {\em Operations Research}, 16(5):955--961.

\bibitem[Vert and Bleakley, 2010]{NIPS2010_4157}
Vert, J. and Bleakley, K. (2010).
\newblock Fast detection of multiple change-points shared by many signals using
  group {LARS}.
\newblock In Lafferty, J., Williams, C., Shawe-Taylor, J., Zemel, R., and
  Culotta, A., editors, {\em Advances in Neural Information Processing Systems
  23}, pages 2343--2351. Curran Associates, Inc.

\bibitem[Wang and Samworth, 2017]{wang2016high}
Wang, T. and Samworth, R.~J. (2017).
\newblock High-dimensional changepoint estimation via sparse projection.
\newblock {\em Journal of the Royal Statistical Society, Series B
  (Forthcoming)}.

\bibitem[Wooldridge, 2010]{panel_data}
Wooldridge, J.~M. (2010).
\newblock {\em Econometric analysis of cross section and panel data}.
\newblock MIT Press.

\bibitem[Xie and Siegmund, 2013]{xie2013}
Xie, Y. and Siegmund, D. (2013).
\newblock Sequential multi-sensor change-point detection.
\newblock {\em Annals of Statistics}, 41(2):670--692.

\bibitem[Yao, 1987]{yao1987}
Yao, Y.-C. (1987).
\newblock Approximating the distribution of the maximum likelihood estimate of
  the change-point in a sequence of independent random variables.
\newblock {\em Annals of Statistics}, 15(3):1321--1328.

\bibitem[Zhang and Siegmund, 2007]{Zhang2007}
Zhang, N.~R. and Siegmund, D.~O. (2007).
\newblock {A modified Bayes information criterion with applications to the
  analysis of comparative genomic hybridization data}.
\newblock {\em Biometrics}, 63:22--32.

\end{thebibliography}
\end{document}